# Building a digital twin of EDFA: a grey-box modeling approach

Yichen Liu,[1] Xiaomin Liu,[1] Yihao Zhang,[1] Meng Cai,[1] Mengfan Fu,[1] Xueying Zhong,[1] Lilin Yi,[1] Weisheng Hu,[1,2] and Qunbi Zhuge[1,2,*]

[1]*State Key Laboratory of Advanced Optical Communication Systems and Networks, Department of Electronic Engineering, Shanghai Jiao Tong University, Shanghai 200240, China*
[2]*Peng Cheng Laboratory, Shenzhen 518000, China*
*\*Corresponding author: qunbi.zhuge@sjtu.edu.cn*

**Abstract:** To enable intelligent and self-driving optical networks, high-accuracy physical layer models are required. The dynamic wavelength-dependent gain effects of non-constant-pump erbium-doped fiber amplifiers (EDFAs) remain a crucial problem in terms of modeling, as it determines optical-to-signal noise ratio as well as the magnitude of fiber nonlinearities. Black-box data-driven models have been widely studied, but it requires a large size of data for training and suffers from poor generalizability. In this paper, we derive the gain spectra of EDFAs as a simple univariable linear function, and then based on it we propose a grey-box EDFA gain modeling scheme. Experimental results show that for both automatic gain control (AGC) and automatic power control (APC) EDFAs, our model built with 8 data samples can achieve better performance than the neural network (NN) based model built with 900 data samples, which means the required data size for modeling can be reduced by at least two orders of magnitude. Moreover, in the experiment the proposed model demonstrates superior generalizability to unseen scenarios since it is based on the underlying physics of EDFAs. The results indicate that building a customized digital twin of each EDFA in optical networks become feasible, which is essential especially for next generation multi-band network operations.

## 1. Introduction

To support various emerging Internet applications, such as artificial intelligence, virtual/augmented reality, cloud services and video streaming, optical networks are facing with an explosive growth in capacity demand. Over the past decade, attributed to the rapid advances in modulation and coding [1] as well as digital signal processing (DSP) [2], the capacity of optical transmission systems is now very close to the Shannon limit [3].

In order to further exploit the network capacity, more efficient network control and management is required. In this context, novel network operation paradigms including software-defined network (SDN) [4], elastic optical network (EON) [5], and self-driving optical network have been investigated and developed. These control and management strategies bring out the necessity to acquire accurate and fine models of the physical layer [6,7]. The most significant challenge in constructing such physical layer models lies in evaluating the complex physical processes as signals pass through optical components including fibers, optical amplifiers, wavelength selective switches (WSSs) and optical transceivers [8,9].

Among the physical layer models, the model of erbium-doped fiber amplifier (EDFA) is critical [10,11]. This is because the behavior of EDFAs directly determines the optical signal-to-noise ratio (OSNR) of a transmitted signal [12]. Moreover, the optical power at the output of an EDFA determines the magnitude of fiber nonlinearities [13] in the following fiber span. These are the major impairments in a long-haul optical transmission system. Therefore, the accuracy of EDFA models largely determines whether the quality of transmission can be accurately estimated.

On the other hand, to provide more spectrum resources, multi-band optical transmission systems are being commercialized [14]. In a multi-band system, multiple rare-earth doped fiber amplifiers are utilized to amplify signals on different wave bands [15]. The operating conditions of these amplifiers should be jointly optimized to ensure the transmission performance in the presence of inter-band stimulated Raman scattering (ISRS) [16]. In this scenario, the modeling of fiber amplifiers become even more critical.

We focus on the modeling of EDFA's gain in this paper. The EDFA gain characteristics are complicated to analyze due to the static-state inter-channel gain variation and the dynamic gain excursion. First of all, the gain of EDFA is a non-flat spectrum [17]. Multiple underlying physics such as the homogeneous and inhomogeneous broadening [18] result in a wavelength-dependent gain characteristic. Second, the gain spectrum varies significantly on different input signal configurations and device settings [19]. EDFA often works in an automatic gain control (AGC) or automatic power control (APC) mode to maintain a constant gain or output power. In these working modes, the pump power is dynamically adjusted, leading to a dynamic gain excursion [20,21]. Additionally, the characteristics of various physical devices are usually inconsistent due to the diversity in physical design and variance in fabrication [22]. This further necessitates the requirement for customized modeling of each individual EDFA.

Many investigations on non-constant-pump EDFA models have been reported. First, based on the underlying physics, several explicit EDFA gain models are proposed [17,18,21]. However, these models cannot satisfy the requirement of high accuracy in practical use. In recent years, data-driven EDFA gain models [10,23,24] have attracted increasing attention for their ability to provide accurate estimations. For instance, with a relatively large dataset (about 40,000 data samples in [23] and 12,000 data samples in [10]) for training, neural networks (NNs) are used to estimate the gain or noise power spectrum with a root mean square error (RMSE) below 0.1 dB. Some other NN-based models are further proposed to improve the model's generalizability [22,25], fasten the training process [10,26], reduce the required training dataset [27], or improve the modeling accuracy [28,29]. However, it is difficult to apply these NN-based models in practical systems. This is primarily because that the large dataset required to train these models is typically either unavailable or too costly to obtain in real systems. In addition, the NN-based model is generally a black-box model with poor interpretability and generalizability, making it difficult to guarantee its performance in complex and dynamic systems.

In this work, we aim at providing an easy-to-implement gain model of EDFA by greatly reducing the required data size and improving the model's generalizability. We analyze the physics of a typical class of EDFAs and show that their gain spectra are functions of a single independent variable. Then, based on this physical derivation, we propose an accurate grey-box model, which only needs 8 data samples to customize a digital twin model for each individual EDFA. The performance of this model is verified on both our experimentally measured dataset and a public dataset [30]. Compared with the pure NN-based model, the proposed model reduces the required data size by at least two orders of magnitude, while achieving a better accuracy and generalizability.

The remaining part of the article is structured as follows. In Section 2, the underlying physics of EDFA is introduced, and the EDFA gain characterization is analyzed. In Section 3, the proposed grey-box modeling scheme is described. In Section 4, the experimental setup and dataset are presented. In Section 5, the superior performance of the modeling scheme is demonstrated on both our experimental dataset and the public dataset.

## 2. Analytical analysis of EDFA gain characteristics

The EDFA gain spectrum is a complex function of input signal spectrum, EDFA settings, as well as multiple underlying parameters, which are generally difficult to acquire. In this section, we first describe the EDFA's underlying physics and mathematics by introducing its structure and the process of optical amplification. Afterwards, we derive a simple expression with only one dynamic variable to describe the EDFA gain behaviors. The simplified expression will serve as the basic framework of the proposed grey-box EDFA gain model.

### 2.1 Typical structure of EDFA

The structure of a typical single-stage EDFA [31] is shown in Fig. 1. The key component is an erbium-doped fiber (EDF), in which the signal light is amplified through the process of the stimulated emission transition with pump lights. At both the input and output of the EDF, isolators are adopted to eliminate the effects of reflected lights, such as self-excited oscillation. A gain flattening filter (GFF) is usually used to reduce the wavelength-dependent variation of gain [32]. A control circuit is employed to monitor the input and output power of the EDFA, and then adjust the powers of the pumps. When the EDFA is set to work in an AGC or APC mode, the powers of the pumps are dynamically adjusted to maintain a total gain for AGC or total output power for APC, respectively.

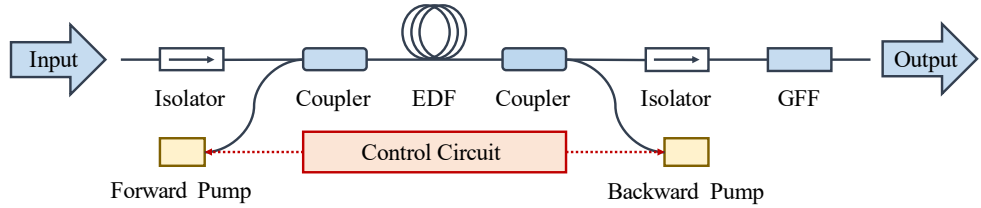

Fig. 1. Structure of a typical single-stage EDFA.

To achieve a larger gain and/or a lower NF, a multi-stage structure of EDFA is often commercially adopted [33]. In such a structure, EDFs cascade and each stage may contain GFFs and optical attenuators.

In the following, the gain characteristics of the multi-stage EDFAs under different input signals and amplifier settings are analyzed. When the input signal and amplifier setting vary, most of the internal structures such as isolator, GFF, and coupler, are generally considered to provide a static response. The EDFA gain variation is mainly caused by the dynamic pump power adjustment and the EDF response variation with different signals and pumps. Next, we will introduce the analytical expression of the EDF's response.

### 2.2 Gain characterization of EDF

A widely applied model of Saleh *et al*. [17] accurately predicts the evolution of signal power along the EDF, which can be expressed as

$$\frac{dP_\lambda(z)}{dz \cdot P_\lambda(z)} = \rho \Gamma_\lambda [\sigma_{e,\lambda} N_2(z) - \sigma_{a,\lambda} N_1(z)] - \alpha_s, \tag{1}$$

where $\rho$ is the $Er^{3+}$ ions concentration and $\alpha_s$ is the background loss of the signal. $\sigma_{a,\lambda}$ and $\sigma_{e,\lambda}$ are the absorption and emission cross-sections at wavelength $\lambda$, respectively. $\Gamma_\lambda$ is the signal overlap factor at wavelength $\lambda$. $N_1(z)$ and $N_2(z)$ are the lower and upper level's fractional population at a fiber length $z$, respectively, which are both related to the signal and pump power at a fiber length $z$ [17]. For a two-level EDFA [18], only the two energy levels of the $Er^{3+}$ ions are used, which means $N_1(z) + N_2(z) = 1$.

By substituting the relationship $N_1(z) = 1 - N_2(z)$ into Equation (1) and integrating both sides of the equation, the gain on a logarithmic scale can be obtained as

$$\ln[P_\lambda(L)] - \ln[P_\lambda(0)] = \rho\Gamma_\lambda[(\sigma_{e,\lambda} + \sigma_{a,\lambda})] \cdot \langle N_2\rangle - (\rho\Gamma_\lambda \cdot \sigma_{a,\lambda} + \alpha_s) \cdot L, \tag{2}$$

where $\langle N_2\rangle$ is the integration of $N_2(z)$ over a fiber length $L$: $\langle N_2\rangle = \int_0^L N_2(z)\,dz$.

In Equation (2), only the value of term $\langle N_2\rangle$ is related to the signals and pumps [34-37]. All the other terms are constant parameters of the EDF. By combining the constant terms, a simplified expression can be derived as

$$\boldsymbol{G}_{EDF}(L) = \boldsymbol{A} \cdot \langle N_2\rangle + \boldsymbol{B} \cdot L, \tag{3}$$

where $\boldsymbol{G}_{EDF}(L)$ is the EDF's wavelength-dependent gain spectrum on the dB scale, and its value at wavelength $\lambda$ is: $G_{EDF}(L,\lambda) = 4.343 \cdot \ln[P_\lambda(L)/P_\lambda(0)]$. $\boldsymbol{A}$ and $\boldsymbol{B}$ are constant wavelength-dependent spectra for a certain type of EDF, and their values at wavelength $\lambda$ are: $A(\lambda) = 4.343 \cdot \rho\Gamma_\lambda[(\sigma_{e,\lambda} + \sigma_{a,\lambda})]$, and $B(\lambda) = -4.343 \cdot (\rho\Gamma_\lambda \cdot \sigma_{a,\lambda} + \alpha_s)$. Equation (3) indicates that for arbitrary signals and pumps, the logarithmic gain of an EDF experienced by each wavelength is only linearly related to a single variable $\langle N_2\rangle$. Note that throughout this paper, the gain spectrum will appear on the dB scale.

*2.3 Gain characterization of EDFA*

Next, the gain spectrum of a complete multi-stage EDFA is derived considering all the internal components of EDFA. The response of the *i*th stage of GFFs, couplers, etc., is regarded as a constant insertion loss spectrum, represented by $\boldsymbol{Loss}_i$. The overall gain spectrum of an *n*-stage EDFA can be expressed as

$$\boldsymbol{G}_{overall} = \sum_{i=1}^{n}(\boldsymbol{A}_i \cdot \langle N_2\rangle_i + \boldsymbol{B}_i \cdot L_i + \boldsymbol{Loss}_i). \tag{4}$$

For a certain EDFA, the EDFs of each stage are commonly of the same specification. In this case, the parameters $\boldsymbol{A}_i$ and $\boldsymbol{B}_i$ are the same constants for each stage, denoted as $\boldsymbol{A}_{const}$ and $\boldsymbol{B}_{const}$, respectively. And then the equation can be simplified to

$$\boldsymbol{G}_{overall} = \boldsymbol{A}_{const} \cdot k + \boldsymbol{C}_{const}, \tag{5}$$

where $\boldsymbol{C}_{const} = \boldsymbol{B}_{const} \cdot \sum_{i=1}^{n} L_i + \sum_{i=1}^{n} \boldsymbol{Loss}_i$, $k = \sum_{i=1}^{n}\langle N_2\rangle_i$. It should be noted that there may be a variable optical attenuator (VOA) between each pair of EDFs, but we assume it holds a constant attenuation when the EDFA is operated under a given gain/output power setting. In this case, $\boldsymbol{A}_{const}$ and $\boldsymbol{C}_{const}$ are constant spectra, and $k$ is a dynamic variable related to input signals. Equation (5) holds under arbitrary input signals and pump configurations, indicating that for a certain *n*-stage EDFA, the gain spectrum is a univariate linear function.

On the other hand, for AGC/APC EDFAs, the control circuit dynamically adjusts the pump power to meet the gain/power settings. The relationship of the input signal, the gain/power setting, and the corresponding gain spectrum is

$$\begin{cases} \sum 10^{(G_{overall}+P_{in})/10} = \widehat{G_{set}} + P_{in,total}, & \text{(for AGC mode)} \\ \sum 10^{(G_{overall}+P_{in})/10} = \widehat{P_{set}}, & \text{(for APC mode)} \end{cases} \tag{6}$$

where $\boldsymbol{P}_{in}$ denotes the input power spectrum. $P_{in,total}$ is the total power of the input signal. Due to the inaccuracy of the power monitoring process, there may be a difference between the target gain/power and the real gain/power, which can be calibrated.

### 3. Grey-box EDFA gain model

The above mathematical framework proves that for an EDFA in which the EDFs are of the same specification, the gain spectrum is a univariate linear function, and the total gain/power is related to the AGC/APC setting. Based on this inference, we propose an accurate grey-box EDFA gain model based on a minuscule dataset. The steps of the modeling scheme are as follows: First, a minuscule dataset of EDFA gain spectrum is collected. Second, the gain spectra are written as a univariate linear function: $\boldsymbol{G}_{overall} = \Delta\boldsymbol{G} \cdot x + \boldsymbol{G}_0$, which will be explained in detail in the following part of this section. The parameters in the function, namely ($\Delta G$, $G_0$), can be easily determined through simple processing of the dataset by calculating the variation and average of the gain spectra. Third, the variable $x$ in the linear function is determined based on Equation (6), which will also be detailed later. Finally, with the results of the second and third steps, the EDFA gain spectrum can be constructed. The framework of the modeling scheme is shown in Fig. 2. In the following, the detailed procedures of the second and third steps are respectively described.

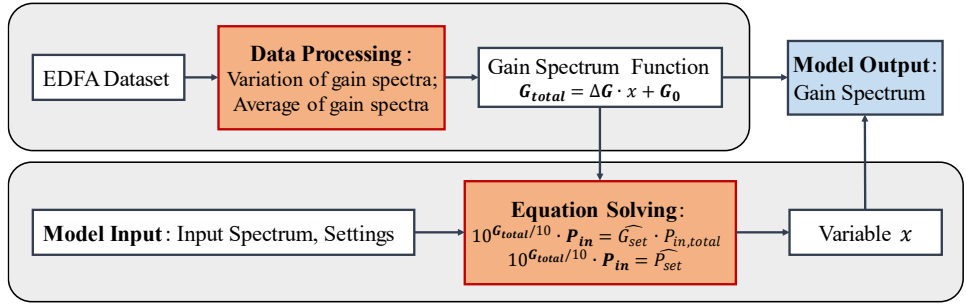

Fig. 2. Proposed grey-box EDFA gain modeling scheme.

### 3.1 Determination of the gain spectrum function

In the second step, a simple data processing method is proposed to determine the univariable equation of the EDFA gain spectrum. Equation (5) presents the relationship between the gain spectrum and a single variable $k$. However, the equation is difficult to obtain for the unavailability of parameters, but it can be equivalently transformed to other expressions. Let a variable $x$ be an arbitrary linear transformation to the variable $k$, which is: $x = a \cdot k + b$, where $a$, $b$ are arbitrary constants except for $a = 0$. Equation (5) can be rewritten as

$$G_{overall} = \Delta G \cdot x + G_0, \tag{7}$$

where $\Delta G$, $G_0$ are constant spectra, respectively denoted as: $G_0 = -A_{const} \cdot b/a + C_{const}$, and $\Delta G = A_{const}/a$.

A set of parameters ($\Delta G$, $G_0$) can be determined by using a measured EDFA gain spectrum dataset. A small dataset containing $n$ EDFA gain spectra is given as $G_{1 \sim n}$. Each spectrum in the dataset corresponds to a value of the variable $k$, and the gain spectrum dataset corresponds to a set of $n$ values of the variable $k$. There must exist a linear transformation to transform the variable $k$ into a variable $x$ such that the $n$ values of $x$ have an average of 0 and a range of 1. Each gain spectrum in the dataset can be expressed as Equation (7), and a total of $n$ equations with different values of the variable $x$ can be set up. As the distribution of the values of variable $x$ is known, we can solve the equations to obtain the parameters ($\Delta G$, $G_0$). $\Delta G$ can be regarded as the variation of the gain spectra, which is solved by taking the difference of the highest and lowest gain spectra's corresponding equations. As the range of the values of $x$ is 1, $\Delta G$ can be represented as

$$\Delta G = G_{max} - G_{min}, \tag{8}$$

where $G_{max}$ and $G_{min}$ are the gain spectra with the highest and lowest total gain, respectively. To avoid random errors and provide a more accurate result, $G_{max}$ and $G_{min}$ can be replaced by the average of several gain spectra with the largest and smallest total gains. $G_0$ can be regarded as the average of the gain spectra, which is solved by calculating the average of all the $n$ gain spectra's corresponding equations. As the average of the values of $x$ is 0, $G_0$ can be represented as

$$G_0 = \frac{1}{n} \sum_{i=1}^{n} G_i. \tag{9}$$

*3.2 Estimation of variable*

In the third step, on determining the univariate linear function of the gain spectrum, we only need to estimate the variable $x$ to obtain the model. Equation (6) provides the relationship of the input signal, gain/power setting and the corresponding gain spectrum, which can be substituted into Equation (6), showing as a function of variable $x$:

$$\begin{aligned}\sum 10^{(\Delta G \cdot x + G_0 + P_{in})/10} &= \widehat{G_{set}} + P_{in,total}, &\text{(for AGC mode)} \\ \sum 10^{(\Delta G \cdot x + G_0 + P_{in})/10} &= \widehat{P_{set}}. &\text{(for APC mode)}\end{aligned} \quad (10)$$

This is a monotone equation about the variable $x$. Therefore, the value of the variable can be uniquely determined.

Finally, based on the gain spectrum function determined in the second step and the value of the variable estimated in the third step, the gain spectrum can be constructed by Equation (7).

## 4. Datasets for modeling performance evaluation

The performance of the proposed modeling scheme is evaluated on both our experimentally measured dataset and a public dataset. In this section, we first introduce our experimental setup and the measured dataset, and then briefly introduce the public dataset.

*4.1 Experimental setup and measured dataset*

The experimental system for data collection is shown in Fig. 3 (a). An amplified spontaneous emission (ASE) source is utilized to generate a wideband signal. The wideband signal is then spectrally shaped by a programmable filter to generate signals with different channel loading conditions. The filtered signal, ranging in frequency from 192.1 THz to 196.1 THz, is composed of 80 channels with 50 GHz bandwidth. Two optical spectrum analyzers (OSAs) are utilized to measure the input and output spectra of the EDFA under test (EUT). For our experiment, the EUT is set to work in APC mode with a target output power of 15 dBm and AGC mode with a target gain of 18 dB.

The total 80 channels are divided into odd-number channels and even-number channels. The even-number channels are set to empty to measure ASE noise. The odd-number channels are set to be loaded or unloaded in a random manner. Besides, a power perturbation is applied to the loaded channels. The average power of the loaded channels in an input spectrum ranges from about -18 dBm to about -14 dBm. The power of the unloaded channels is about -28 dBm. The power perturbations are randomly generated for each loaded channel with a range of -1.5 dB to 1.5 dB. For each setting mode, a total of 1500 data samples are measured. The configuration of the dataset is summarized in Table 1.

**Table 1. Configuration of Our Experimental Dataset**

| Parameters | Value |
|---|---|
| EDFA setting mode | APC, AGC |
| APC power setting (dBm) | 15 |
| AGC gain setting (dB) | 18 |
| Input spectrum shape | Even-number channel: empty<br>Odd-number channel: load/unload randomly |
| Average input power per channel (dBm) | Loaded channels: about -18: -14<br>Unloaded channels: about -28 |
| Power perturbation per channel (dB) | -1.5:1.5 |
| The data size for each setting | 1500 |

In this experiment, to obtain an accurate gain spectrum, the ASE noise generated by the EUT should be excluded from the measured output power spectrum. It is measured based on the empty even-number channels and then subtracted from the odd-number channels. The ASE noise power spectrum on the total 80 channels can be obtained by the interpolation of the output on the even-number channels. Subsequently, the gain spectrum on odd-number channels can be obtained by: $Gain = (P_{out} - P_{noise})/P_{in}$, where $P_{out}$, $P_{noise}$ and $P_{in}$ indicates the output power spectrum, noise power spectrum and input power spectrum, respectively.

The data collection is implemented using an automatic controller with the following three steps: (1) Set the target gain/output power of the EUT and the output power of the ASE source before the experiment. (2) Set the programmable filter to generate the input signal power spectrum. (3) Read and store the data from the two OSAs after the system reach a stable status. These steps are repeated until the whole target input signal spectra are all measured. The control logic is shown in Fig. 3(b).

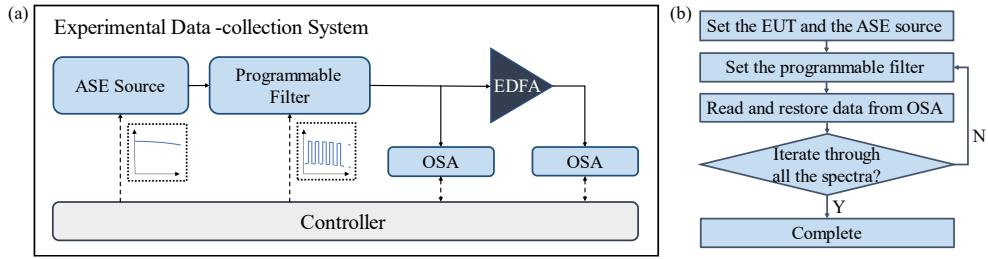

Fig. 3. Experimental setup: (a) the experimental system, (b) the control flow of the automatic controller.

For each AGC/APC setting, there is an inaccuracy in the process of EDFA's power monitoring, and the difference between the target gain/power and the real gain/power should be calibrated.

*4.2 Public dataset*

The public dataset [30] contains datasets of three EDFAs of the same make. The EDFAs are set to work in APC mode. A total of 4 output power settings are considered, covering the range of target output power from 15 to 18 dBm with a step size of 1 dB. Besides, a total of 9 input power conditions are considered, covering the range of total input power from about -6.5 to 1.5 dBm. The input signal is a random continuous spectrum, in which the power of each channel is generated to be close to the power of adjacent channels. In each output power setting and input power scenario, about 2000 sets of input spectra and output spectra are measured. In this paper, the dataset of one EDFA under 15 dBm target output power containing 16497 data samples is investigated. The configuration of the public dataset is summarized in Table 2. The gain spectrum is directly calculated from the output spectrum and input spectrum by $Gain = P_{out} - P_{in}$. It is worth mentioning that the presence of ASE noise can lead to an inaccuracy in the calculation of the gain spectrum. However, considering the fact that the typical ASE noise power is usually much lower than the amplified signal power, the degree of inaccuracy will be acceptable.

Table 2. Configuration of the Adopted Data from the Public Dataset

| Parameters | Value |
|---|---|
| EDFA setting mode | APC |
| APC power setting (dBm) | 15 |
| Input spectrum shape | Continuous power spectrum |
| Total input power (dBm) | About -6.5:1:1.5 |
| The data size for each setting | About 2000 |

## 5. Experimental results and discussions

The performance of the proposed grey-box model for both AGC EDFA and APC EDFA is evaluated. For comparison, the performance of the traditional NN modeling scheme is also shown. The configuration of the NN is similar to the one proposed in [22]. The input of the NN model is the input power spectrum, the total input power and the EUT setting, and the output of the NN model is the gain spectrum. The NN model has two hidden layers. For our dataset with 40 channels, the neuron numbers of the two hidden layers are 128 and 64, respectively. For the public dataset with 83 channels, they are 256 and 128, respectively.

The modeling accuracy is evaluated by the RMSE of the predicted gain spectrum, and 400 data samples are used as the testing dataset.

### 5.1 Modeling performance for AGC EDFA

First, for AGC EDFA, RMSEs of the proposed model and the NN model utilizing different training data sizes are plotted in Fig. 4(a). Multiple rounds of modeling are conducted, in which the training data samples are randomly selected from the dataset, and the obtained fluctuations of RMSEs are shown as the error bars. The result shows that the proposed modeling scheme achieves an average RMSE of 0.062 dB using only 8 data samples. The performance is stable in the multiple rounds of modeling, as the maximum RMSE is 0.070 dB. On the contrary, the NN model requires 900 training data samples to reach an average RMSE of 0.109 dB. We can see that the proposed scheme requires only 0.9% training data samples compared to the NN scheme, while still achieving much better accuracy. The error distribution of the proposed scheme utilizing 8 data samples and the traditional NN scheme trained on 900 data samples is compared in Fig. 4(b). Based on it, the cumulative distribution function (CDF) for the absolute value of error is calculated. When the CDF reaches 90%, the absolute value of error is 0.104 dB for the proposed modeling scheme, while it is 0.219 dB for the traditional NN modeling scheme.

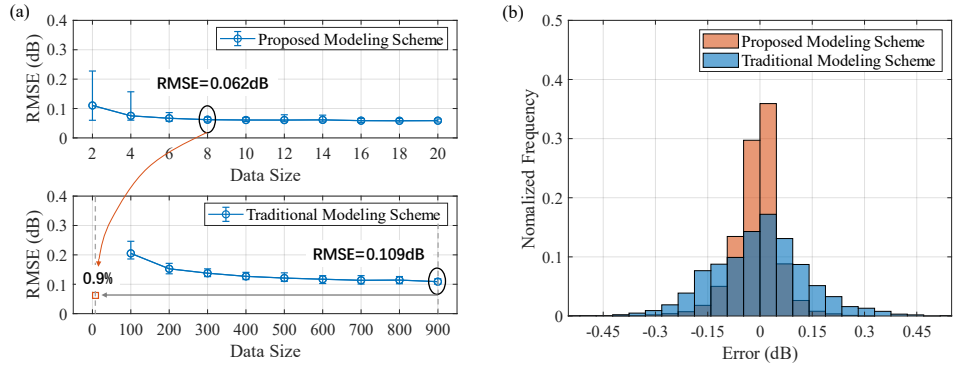

Fig. 4. Modeling performance for AGC EDFA on our experimental dataset: (a) the modeling accuracy versus the utilized data size, (b) the distribution of the modeling error.

Next, the proposed model's superior generalizability is demonstrated. The model is established on the data samples with more than 12 loaded channels, and then the performance is verified on the data samples with less than 12 loaded channels. The RMSE of the proposed model utilizing 8 data samples reaches 0.087 dB, whereas the RMSE of the traditional NN model utilizing 900 data samples is 0.262 dB. The comparison of error distribution is shown in Fig. 5. The absolute value of error is 0.149 dB for the proposed modeling scheme and 0.522 dB for the traditional NN modeling scheme when the CDF reaches 90%. This result illustrates that the black-box NN model suffers from poor generalizability, and the performance of the model is strongly related to the distribution of the input dataset. An NN model should be retrained on another large dataset when utilized in new scenarios. On the contrary, the proposed grey-box model benefits from the knowledge of the underlying physics, and can be easily generalized to other scenarios.

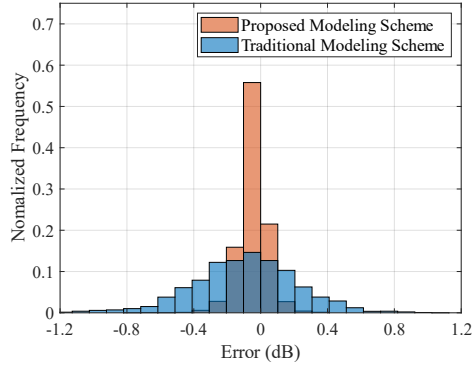

Fig. 5. Generalizability performance for AGC EDFA on our experimental dataset.

### 5.2 Modeling performance for APC EDFA

First, the data efficiency of the proposed modeling scheme for APC EDFA is demonstrated on our experimentally measured dataset. RMSEs of the proposed model and the NN model utilizing different data sizes are plotted in Fig. 6(a). The result shows that the proposed modeling scheme achieves an average RMSE result of 0.070 dB when the model is established using only 8 data samples. The performance is stable in multiple rounds of modeling, as the maximum RMSE is 0.085 dB. On the contrary, the NN model requires 900 training data samples to reach an average RMSE of 0.112 dB. The proposed scheme requires only 0.9% training data samples compared to the NN scheme, while achieving better accuracy. The error distribution of the proposed scheme utilizing 8 data samples and the traditional NN scheme trained on 900 data samples is compared in Fig. 6(b). The absolute value of error is 0.109 dB for the proposed modeling scheme and 0.247 dB for the traditional NN modeling scheme when the CDF reaches 90%.

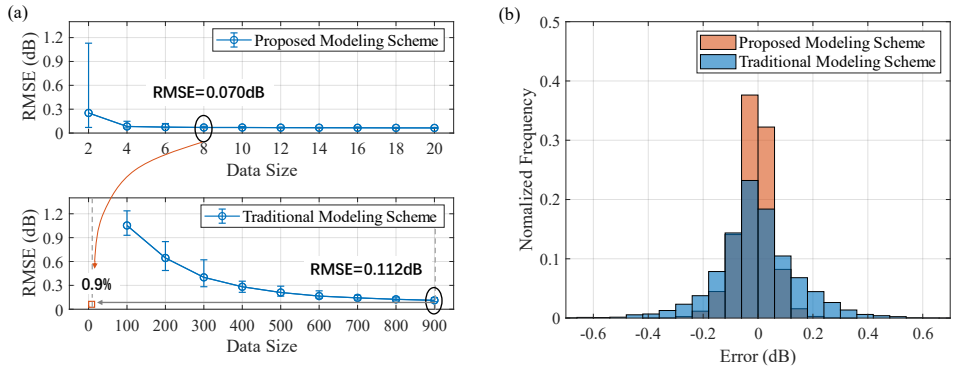

Fig. 6. Modeling performance for APC EDFA on our experimental dataset: (a) the modeling accuracy versus the utilized data size, (b) the distribution of the modeling error.

The evaluation of the model's generalizability is conducted with the same channel loading conditions as the AGC EDFA. The RMSE of the proposed model with 8 data samples reaches 0.093 dB, whereas the RMSE of the NN model with 900 data samples is 0.438 dB. The error distributions are shown in Fig. 7. For a CDF of 90%, the absolute value of error is 0.157 dB and 0.772 dB for the two models, respectively. For the NN model, an offset of about 0.3 dB exists in the results. The reason is that besides the channel loading condition, the APC EDFA's pump adjustment process is also sensitive to the total input power. The data utilized to establish the model and the data utilized to verify the performance are totally different in the distribution of total input power. For the NN model, the difference in distribution is not considered, leading to poor generalization performance. For the proposed grey-box model, the difference in distribution is taken into account by the embedded physics, so the performance is not seriously deteriorated.

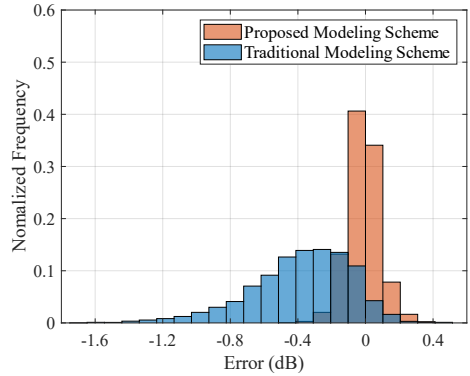

Fig. 7. Generalizability of APC EDFA model on our experimental dataset.

For the public dataset, similar results are observed. As shown in Fig. 8(a), our model with 8 data achieves an average RMSE result of 0.099 dB and a maximum RMSE result of 0.112 dB, whereas the NN model with 900 data achieves an average RMSE result of 0.120 dB and a maximum RMSE result of 0.253 dB. Fig. 8(b) plots the error distributions. For a CDF of 90%, the absolute value of error is 0.186 dB and 0.330 dB for the two models, respectively.

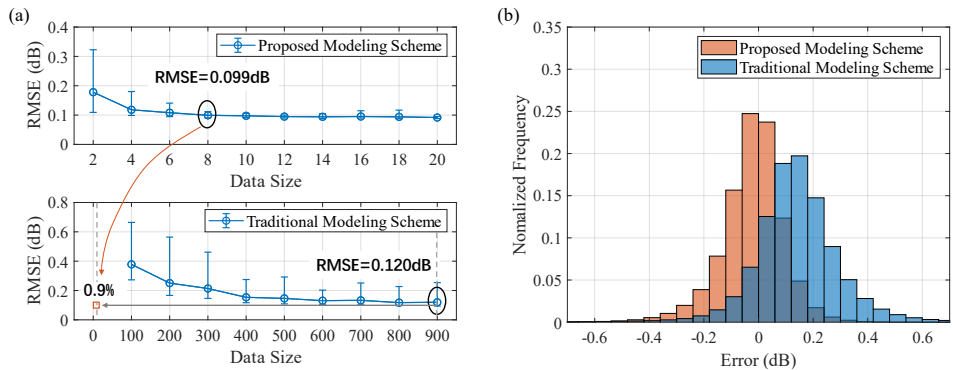

Fig. 8. Modeling performance for APC EDFA on the public dataset: (a) the modeling accuracy versus the utilized data size, (b) the distribution of the modeling error.

For the evaluation of the generalizability, the models are established on the data samples with a total input power ranging from -6.5 dBm to -4.5 dBm, and then the performance is verified on the data samples with a total input power ranging from -3.5 dBm to 1.5 dBm. The RMSE is 0.163 dB and 4.825 dB for the two models, respectively. Fig. 9 plots the error distributions. The results demonstrate that the proposed modeling scheme performs well on different physical devices.

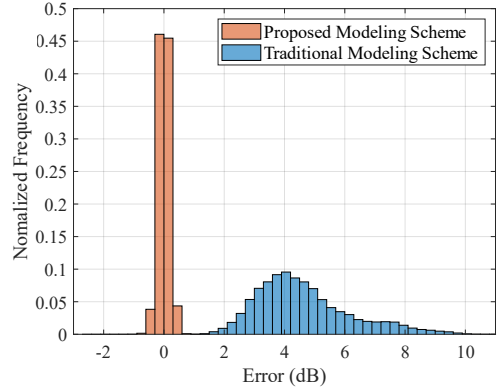

Fig. 9. Generalizability of APC EDFA model on the public dataset.

## 6. Conclusion

A grey-box EDFA gain modeling scheme is proposed to provide accurate estimation based on small datasets. We first analyze the typical structure of EDFA and the physics of optical amplification process, and then derive the EDFA gain spectrum as a univariable linear function. Then, based on the principles of the control circuit's dynamic pump adjustment, the function of the gain spectrum can be uniquely solved, and the gain spectrum can be further obtained. Experiments are carried out to verify the modeling performance. The proposed scheme is proved to be capable of constructing a model using only 8 data samples, while still outperforming the traditional NN model with 900 data samples on accuracy and generalizability. In addition, the performance is also verified in a public dataset, and similar results are achieved. This implies that the modeling scheme is applicable to a variety of physical devices and can be used to build a digital twin of each EDFA in optical networks.

**Funding.** The Shanghai Pilot Program for Basic Research - Shanghai Jiao Tong University (21TQ1400213), and National Natural Science Foundation of China (62175145).

**Disclosures.** The authors declare no conflict of interest.

**Data availability.** Data underlying the results presented in this paper are not publicly available at this time but may be obtained from the authors upon reasonable request.

**References**

1. P. J. Winzer, D. T. Neilson, and A. R. Chraplyvy, "Fiber-optic transmission and networking: the previous 20 and the next 20 years [Invited]," Opt. Express **26**, 24190-24239 (2018).
2. S. J. Savory, "Digital coherent optical receivers: algorithms and subsystems," IEEE J. Sel. Top. Quantum Electron. **16**, 1164–1179 (2010).
3. K. Roberts, Q. Zhuge, I. Monga, S. Gareau, and C. Laperle, "Beyond 100 Gb/s: capacity, flexibility, and network optimization," J. Opt. Commun. Netw. **9**, C12 (2017).
4. P. N. Ji, "Software defined optical network," in *International Conference on Optical Communications and Networks (ICOCN)* (2012).
5. O. Gerstel, M. Jinno, A. Lord, and S. J. B. Yoo, "Elastic optical networking: a new dawn for the optical layer?," IEEE Commun. Mag. **50**, s12–s20 (2012).
6. W. Mo, C. L. Gutterman, Y. Li, S. Zhu, G. Zussman, and D. C. Kilper, "Deep-neural-network-based wavelength selection and switching in ROADM systems," J. Opt. Commun. Netw. **10**, D1 (2018).

7. E. Seve, J. Pesic, and Y. Pointurier, "Accurate QoT estimation by means of a reduction of EDFA characteristics uncertainties with machine learning," in *International Conference on Optical Network Design and Modeling (ONDM)* (2020).
8. A. Minakhmetov, B. Prieur, M. L. Monnier, D. Rouvillain, and B. Lavigne, "Digital twin of unrepeatered line based on Raman and remote optically pumped amplifier machine learning models," in *Optical Fiber Communications Conference (OFC)* (2023), p. W4H.4.
9. Q. Zhuge, "AI-driven digital twin for optical networks," in *European Conference on Optical Communication (ECOC)* (2022), p. Mo3A.1.
10. J. Yu, S. Zhu, C. L. Gutterman, G. Zussman, and D. C. Kilper, "Machine-learning-based EDFA gain estimation [Invited]," J. Opt. Commun. Netw. **13**, B83 (2021).
11. Y. Liu, X. Liu, L. Liu, Y. Zhang, M. Cai, L. Yi, W. Hu, and Q. Zhuge, "Modeling EDFA gain: approaches and challenges," Photonics **8**, 417 (2021).
12. Y. You, Z. Jiang, and C. Janz, "OSNR prediction using machine learning-based EDFA models," in *European Conference on Optical Communication (ECOC)* (2019).
13. Q. Zhuge, X. Zeng, H. Lun, M. Cai, X. Liu, L. Yi, and W. Hu, "Application of machine learning in fiber nonlinearity modeling and monitoring for elastic optical networks," J. Light. Technol. **37**, 3055–3063 (2019).
14. A. Ferrari, A. Napoli, J. K. Fischer, N. Costa, A. D'Amico, J. Pedro, W. Forysiak, E. Pincemin, A. Lord, A. Stavdas, J. P. F.-P. Gimenez, G. Roelkens, N. Calabretta, S. Abrate, B. Sommerkorn-Krombholz, and V. Curri, "Assessment on the achievable throughput of multi-band ITU-T G.652.D fiber transmission systems," J. Light. Technol. **38**, 4279–4291 (2020).
15. L. Rapp and M. Eiselt, "Optical amplifiers for multi–band optical transmission systems," J. Light. Technol. **40**, 1579–1589 (2022).
16. M. Ibrahimi, G. S. Sticca, F. Musumeci, A. Castoldi, R. Pastorelli, and M. Tornatore, "Selective hybrid EDFA/Raman amplifier placement to avoid lightpath degradation in (C+L) Networks," in *European Conference on Optical Communication (ECOC)* (2022).
17. A. A. M. Saleh, R. M. Jopson, J. D. Evankow, and J. Aspell, "Modeling of gain in erbium-doped fiber amplifiers," IEEE Photonics Technol. Lett. **2**, 714–717 (1990).
18. C. R. Giles and E. Desurvire, "Modeling erbium-doped fiber amplifiers," J. Light. Technol. **9**, 271–283 (1991).
19. J. Junio, D. C. Kilper, and V. W. S. Chan, "Channel power excursions from single-step channel provisioning," J. Opt. Commun. Netw. **4**, A1 (2012).
20. M. Hashimoto, M. Yoshida, and H. Tanaka, "The characteristics of WDM systems with hybrid AGC EDFA in the photonics network," in *Optical Fiber Communications Conference (OFC)* (2002), pp. 517–518.
21. K. Ishii, J. Kurumida, and S. Namiki, "Experimental investigation of gain offset behavior of feedforward-controlled WDM AGC EDFA under various dynamic wavelength allocations," IEEE Photonics J. **8**, 1–13 (2016).
22. F. da Ros, U. C. de Moura, and M. P. Yankov, "Machine learning-based EDFA gain model generalizable to multiple physical devices," in *European Conference on Optical Communication (ECOC)* (2020).
23. Y. You, Z. Jiang, and C. Janz, "Machine learning-based EDFA gain model," in *European Conference on Optical Communication (ECOC)* (2018).
24. S. Zhu, C. L. Gutterman, W. Mo, Y. Li, G. Zussman, and D. C. Kilper, "Machine learning based prediction of erbium-doped fiber WDM line amplifier gain spectra," in *European Conference on Optical Communication (ECOC)* (2018).
25. Z. Wang, D. Kilper, and T. Chen, "Transfer learning-based ROADM EDFA wavelength dependent gain prediction using minimized data collection," in *Optical Fiber Communication Conference (OFC)* (2023), p. Th2A.1
26. S. Zhu, C. Gutterman, A. D. Montiel, J. Yu, M. Ruffini, G. Zussman, and D. Kilper, "Hybrid machine learning EDFA model," in *Optical Fiber Communications Conference (OFC)* (2020), p. T4B.4.
27. J. Lin, X. Lin, and Z. Jiang, "Auxiliary neural network assisted machine learning EDFA gain model," in *Optical Fiber Communications Conference (OFC)* (2023), p. M2E.2.
28. Z. Jiang, J. Lin, and H. Hu, "Machine learning based EDFA channel in-band gain ripple modeling," in *Optical Fiber Communications Conference (OFC)* (2022), p. W4I.2.
29. X. Liu, Y. Chen, Y. Zhang, Y. Liu, L. Yi, W. Hu, and Q. Zhuge, "Physics-informed EDFA gain model based on active learning," arXiv:2206.06077 (2022).
30. M. P. Yankov and F. da Ros, "Input-output power spectral densities for three C-band EDFAs and four multi-span inline EDFAd fiber optic systems of different lengths,"(2020).
31. B. Pedersen, M. L. Dakss, B. A. Thompson, W. J. Miniscalco, T. Wei, and L. J. Andrews, "Experimental and theoretical analysis of efficient erbium-doped fiber power amplifiers," IEEE Photonics Technol. Lett. **3**, 1085–1087 (1991).
32. R. Sommer, R. M. Fortenberry, B. Flintham, and P. C. Johnson, "Multiple filter functions integrated into multi-port GFF components," in *Optical Fiber Communication and the National Fiber Optic Engineers Conference (OFC/NFOEC)* (2007), p. JWA24.
33. B. Belmahdi and K. Mazighi, "Implementation efficiency of a two-stage EDFA amplifier with an inverted trapezoidal filter on gain flattening," in *International Conference on Image and Signal Processing and Their Applications (ISPA)* (2022).

34. A. C. Meseguer, J.-C. Antona, A. Bononi, J. Cho, S. Grubb, P. Pecci, O. Courtois, and V. Letellier, "Highly accurate measurement-based gain model for constant-pump EDFA for non-flat WDM inputs," in *Optical Fiber Communication Conference (OFC)* (2021), p. M5C.4.
35. V. M. C. Mathias, A. Carbo Meseguer, J.-C. Antona, A. Bononi, S. Grubb, O. Courtois, and V. Letellier, "Extension of the measurement-based gain model for non-flat WDM inputs and various pump currents," in *European Conference on Optical Communication (ECOC)* (2021).
36. Y. Li, M. Zhang, and M. Tang, "A data-effective black-box EDFA gain model with singular value decomposition," in *Asia Communications and Photonics Conference (ACP)* (2022), pp. 985–987.
37. M. Zhang, Y. Li, W. Li, J. Chen, Y. Chen, and M. Tang, "Data-effective and accurate EDFA gain prediction black-box model," in *Asia Communications and Photonics Conference (ACP)* (2021), p. T4A.98.